# Tendencias longitudinales en la preparación preuniversitaria: una evaluación cohortal a partir de cursos introductorios de Matemática y Física en educación superior (1980–2019)

# Longitudinal Trends in Pre-University Preparation: A Cohort Evaluation Using Introductory Mathematics and Physics Courses (1980–2019)

Hugo Roger Paz
PhD Professor and Researcher Faculty of Exact Sciences and Technology National University of Tucumán
Email: hpaz@herrera.unt.edu.ar
ORCID: https://orcid.org/0000-0003-1237-7983

**Resumen**

La transición entre la educación secundaria y la educación superior constituye un punto crítico en las trayectorias académicas, especialmente en carreras con fuerte carga en ciencias básicas. En distintos sistemas universitarios se han documentado, de manera recurrente, elevados niveles de reprobación y abandono en asignaturas introductorias, aunque la evidencia disponible suele basarse en estudios transversales o períodos temporales acotados. Este trabajo presenta una evaluación longitudinal de la preparación preuniversitaria a partir del análisis de resultados tempranos en Matemática y Física, consideradas cursos introductorios "sensor" de las exigencias académicas iniciales.

El estudio utiliza registros administrativos completos de una universidad pública argentina y analiza cohortes de ingreso comprendidas entre 1980 y 2019, con cobertura censal y enfoque poblacional. La preparación preuniversitaria se operacionaliza como la compatibilidad a nivel de cohorte entre los antecedentes formativos de los estudiantes y las demandas funcionales de los cursos introductorios, observada a través de los resultados en el primer intento. Se estiman, para cada cohorte y según tipo de institución secundaria de procedencia (pública o privada), la probabilidad de aprobación, la probabilidad de no intento evaluativo y la brecha de desempeño entre ambos sectores.

Los resultados evidencian tendencias de largo plazo consistentes: una disminución gradual de las probabilidades de aprobación temprana, un incremento sostenido de la no participación evaluativa y la persistencia de brechas público–privadas moderadas pero estables. Estos patrones sugieren cambios estructurales en la articulación entre la educación secundaria y la universitaria, más que fluctuaciones coyunturales o efectos individuales. El trabajo contribuye a la literatura sobre evaluación educativa en educación superior aportando evidencia longitudinal inédita para el contexto iberoamericano y proponiendo el uso de cursos introductorios como instrumentos de monitoreo institucional de la preparación académica.

**Palabras clave**

Evaluación educativa; Educación superior; Transición secundaria–universidad; Rendimiento académico; Estudios longitudinales; Matemática; Física.

**Abstract**

The transition from secondary to higher education represents a critical point in academic trajectories, particularly in programmes with a strong emphasis on basic sciences. Across different higher education systems, introductory Mathematics and Physics courses consistently concentrate high rates of early failure and attrition, yet most available evidence relies on cross-sectional analyses or limited time spans. This study presents a longitudinal evaluation of pre-university preparation based on early academic outcomes in Mathematics and Physics, conceptualised as "sensor" courses of initial academic demands.

Using complete administrative records from a large public university in Argentina, the analysis covers entry cohorts from 1980 to 2019 with census-level coverage and a population-based approach. Pre-university preparation is operationally defined as cohort-level compatibility between students' prior educational background and the functional demands of introductory university coursework, observed through first-attempt outcomes. For each cohort and by type of secondary school (public or private), we estimate the probability of course approval, the probability of non-attempt (enrolment without evaluative participation), and the public–private success gap.

The results reveal consistent long-term patterns: a gradual decline in early approval probabilities, a sustained increase in non-attempt behaviour, and the persistence of moderate but stable public–private gaps. These findings point to structural changes in the articulation between secondary education and higher education rather than short-term fluctuations or individual-level effects. The study contributes to the international literature on educational evaluation by providing rare long-horizon longitudinal evidence from an Ibero-American context and by proposing introductory science courses as institutional monitoring tools for pre-university academic preparation.

# 1. Introducción

## 1.1 La transición a la educación superior como problema evaluativo

La expansión de la educación superior en las últimas décadas ha transformado de manera sustantiva los sistemas universitarios, dando lugar a procesos de masificación que ampliaron el acceso pero también modificaron el perfil académico promedio de los estudiantes ingresantes (Trow, 2006). En este contexto, la transición entre la educación secundaria y la universitaria se ha consolidado como un punto crítico en las trayectorias educativas, particularmente en programas con alta demanda en ciencias, tecnología e ingeniería (OECD, 2019).

Diversos estudios internacionales coinciden en señalar que las asignaturas introductorias de Matemática y Física funcionan como "cuellos de botella" académicos, concentrando elevadas tasas de reprobación, rezago y abandono en los primeros años de carrera (Faulkner, Hannigan, & Gill, 2010; Heublein & Wolter, 2011). Esta situación ha sido documentada en sistemas educativos con modelos de acceso selectivo y no selectivo, lo que sugiere que las dificultades observadas no pueden atribuirse exclusivamente a mecanismos de admisión, sino a desajustes más amplios en la articulación entre niveles educativos (Lawson, 2000; OECD, 2020).

En el ámbito anglosajón y europeo, la evidencia longitudinal basada en pruebas diagnósticas ha mostrado una disminución sostenida en la fluidez matemática de los estudiantes que ingresan a carreras científicas y tecnológicas, aun cuando los títulos secundarios mantienen formalmente su validez como credenciales de acceso (Lawson, 2000; Faulkner et al., 2010). En América Latina, y particularmente en países con políticas de acceso irrestricto, la selección tiende a desplazarse hacia el primer año universitario, generando tasas de abandono tempranas especialmente elevadas en carreras de ingeniería y ciencias básicas (CONFEDI, 2018; Secretaría de Políticas Universitarias, 2020).

Desde una perspectiva de evaluación educativa, estas dificultades tempranas plantean un problema central: la falta de indicadores longitudinales que permitan monitorear, de manera sistemática y comparable en el tiempo, el grado de alineación entre la formación previa de los estudiantes y las exigencias académicas iniciales de la educación superior. La mayor parte de los estudios disponibles se concentra en cohortes individuales, intervenciones específicas o análisis transversales, lo que limita la capacidad de identificar tendencias estructurales de largo plazo (OECD, 2019; Heublein & Wolter, 2011).

En este marco, el presente estudio aborda la transición secundaria–universidad como un problema evaluativo y propone el análisis de resultados tempranos en cursos introductorios de Matemática y Física como una estrategia indirecta para evaluar la preparación preuniversitaria a nivel de cohorte. Al utilizar registros administrativos completos durante un período de cuatro décadas, el trabajo busca aportar evidencia empírica robusta sobre la evolución de los patrones de desempeño inicial y contribuir al debate internacional sobre la articulación entre la educación secundaria y la educación superior desde una perspectiva longitudinal y poblacional.

## 2. Marco teórico y estado del arte

### 2.1 Preparación preuniversitaria y compatibilidad académica

La noción de preparación preuniversitaria ha sido abordada en la literatura desde múltiples perspectivas, que incluyen el dominio de contenidos disciplinares, el desarrollo de habilidades cognitivas generales y la adquisición de disposiciones académicas necesarias para el estudio autónomo en contextos universitarios (Kuh et al., 2008; OECD, 2019). Sin embargo, gran parte de estos enfoques se apoyan en mediciones individuales —pruebas diagnósticas, encuestas de autopercepción o evaluaciones estandarizadas— que, si bien resultan útiles para caracterizar perfiles de ingreso, presentan limitaciones para el análisis longitudinal a gran escala y para la comparación intercohortal en períodos extensos.

En respuesta a estas limitaciones, algunos trabajos han propuesto evaluar la preparación preuniversitaria de manera indirecta, a partir del desempeño temprano en asignaturas introductorias consideradas estructuralmente estables en el tiempo (Lawson, 2000; Faulkner et al., 2010). Desde esta perspectiva, la preparación no se concibe como una propiedad individual aislada, sino como una relación de compatibilidad entre los antecedentes formativos promedio de una cohorte y las demandas funcionales de los cursos iniciales de la educación superior. Esta aproximación resulta especialmente pertinente en sistemas con acceso irrestricto o baja selectividad, donde los mecanismos formales de admisión no filtran significativamente por desempeño académico previo (Heublein & Wolter, 2011).

La literatura reciente en evaluación educativa ha destacado la necesidad de indicadores poblacionales que permitan monitorear esta compatibilidad de manera sistemática, evitando interpretaciones normativas o deficitarias sobre los estudiantes (Harackiewicz et al., 2016). En este marco, el análisis de resultados tempranos en cursos introductorios se ha consolidado como una herramienta evaluativa relevante, ya que dichos cursos concentran altas tasas de reprobación y abandono y suelen mantener una estructura curricular relativamente estable a lo largo del tiempo (OECD, 2020).

## 2.2 Cursos introductorios como "sensores" de la transición secundaria–universidad

Las asignaturas iniciales de Matemática y Física ocupan un lugar central en la formación de carreras científicas y tecnológicas y han sido ampliamente estudiadas como puntos críticos de la transición a la educación superior. Investigaciones realizadas en Europa, América del Norte y Oceanía coinciden en que estas materias presentan una fuerte correlación con la permanencia y el avance académico posterior, actuando como filtros tempranos de progresión curricular (Faulkner et al., 2010; Sadler & Tai, 2007).

El concepto de cursos "sensor" se apoya en esta evidencia y refiere a asignaturas cuya posición temprana en el plan de estudios, junto con la estabilidad de sus contenidos y criterios de evaluación, permite utilizarlas como indicadores sensibles de cambios en la preparación académica de las cohortes entrantes (Lawson, 2000). A diferencia de los cursos optativos o avanzados, los cursos sensor suelen ser obligatorios, de cursado masivo y con criterios de aprobación comparables a lo largo del tiempo, lo que los convierte en puntos de observación privilegiados para análisis longitudinales.

Estudios longitudinales basados en este enfoque han documentado, en distintos contextos nacionales, una disminución progresiva en el desempeño promedio en Matemática y Física a lo largo de varias décadas, aun cuando los programas universitarios y los requisitos formales de ingreso se mantuvieron relativamente constantes (Faulkner et al., 2010; Henderson, Sadler, & Sonnert, 2015). Estos hallazgos han sido interpretados como indicios de un desajuste creciente entre la formación secundaria y las exigencias universitarias iniciales, más que como evidencia de un deterioro individual de capacidades cognitivas.

## 2.3 Evidencia internacional sobre tendencias de largo plazo

La revisión de literatura internacional muestra que los problemas asociados a la preparación preuniversitaria no constituyen un fenómeno exclusivamente local o reciente. Informes comparativos de la OECD han señalado que, incluso en sistemas con altos niveles de rendimiento promedio en pruebas estandarizadas, las universidades reportan dificultades crecientes en los conocimientos básicos de los estudiantes ingresantes, particularmente en Matemática (OECD, 2019, 2020).

En el Reino Unido y Australia, análisis basados en pruebas diagnósticas y resultados académicos han evidenciado una disminución sostenida en habilidades matemáticas elementales desde finales del siglo XX, acompañada por un aumento en las estrategias de apoyo y nivelación en el primer año universitario (Lawson, 2000; Henderson et al., 2015). En Estados Unidos, estudios longitudinales han mostrado patrones similares, aunque con fuertes variaciones institucionales y disciplinares (Sadler & Tai, 2007).

En América Latina, la evidencia empírica es más fragmentaria y se concentra mayormente en estudios de caso o análisis de períodos temporales acotados. No obstante, investigaciones recientes basadas en datos administrativos han identificado tasas elevadas y persistentes de reprobación y abandono en los primeros años de carreras científicas y tecnológicas, especialmente en universidades públicas con acceso irrestricto (CONFEDI, 2018; Secretaría de Políticas Universitarias, 2020). La ausencia de series históricas largas ha limitado, hasta el momento, la posibilidad de evaluar si estos patrones responden a fluctuaciones coyunturales o a tendencias estructurales de largo plazo.

## 2.4 Brechas por tipo de institución secundaria

Otro eje recurrente en la literatura es la comparación del desempeño universitario temprano según el tipo de institución secundaria de procedencia. Diversos estudios han documentado brechas persistentes entre estudiantes provenientes de escuelas públicas y privadas, tanto en términos de aprobación inicial como de progresión académica (Harackiewicz et al., 2016; OECD, 2019). Estas diferencias suelen interpretarse como el resultado de desigualdades estructurales en la calidad y los recursos educativos, aunque la magnitud y estabilidad de las brechas varían significativamente entre países y sistemas.

Desde una perspectiva longitudinal, algunos trabajos sugieren que estas brechas tienden a mantenerse relativamente estables en el tiempo, incluso cuando el desempeño promedio de las cohortes se modifica (Faulkner et al., 2010). Este hallazgo resulta particularmente relevante para el análisis de la preparación preuniversitaria, ya que permite distinguir entre cambios globales en la compatibilidad académica y variaciones diferenciales entre subpoblaciones.

En conjunto, la evidencia revisada respalda la pertinencia de abordar la preparación preuniversitaria como un fenómeno estructural y de largo plazo, evaluable mediante indicadores poblacionales derivados del desempeño temprano en cursos introductorios. Sin embargo, la literatura también revela una escasez de estudios con cobertura temporal extensa en contextos iberoamericanos, lo que justifica la contribución empírica del presente trabajo.

**3.1 Diseño del estudio**

El presente trabajo adopta un diseño observacional, longitudinal y retrospectivo, basado en el análisis de datos administrativos universitarios correspondientes a múltiples cohortes de ingreso. El enfoque metodológico se inscribe en la tradición de la evaluación educativa poblacional, orientada a identificar patrones estructurales de desempeño académico temprano a lo largo del tiempo, más que a explicar resultados individuales (OECD, 2020; Harackiewicz et al., 2016).

El estudio se centra en el desempeño en asignaturas introductorias de Matemática y Física, consideradas cursos "sensor" de la transición secundaria–universidad, y analiza su evolución intercohortal durante un período prolongado. Este enfoque permite examinar tendencias de largo plazo en la compatibilidad académica entre la formación secundaria y las exigencias iniciales de la educación superior, evitando sesgos asociados a evaluaciones puntuales o a cohortes aisladas (Lawson, 2000; Faulkner et al., 2010).

## 3.2 Contexto institucional y población de estudio

La investigación se desarrolla en una universidad pública argentina de gran escala, caracterizada por un sistema de acceso irrestricto y por la ausencia de mecanismos selectivos formales de admisión. Este rasgo institucional resulta metodológicamente relevante, ya que permite observar con mayor fidelidad las variaciones en la preparación preuniversitaria promedio de las cohortes entrantes, sin la interferencia de filtros académicos ex ante (Heublein & Wolter, 2011).

La población de estudio está compuesta por estudiantes ingresantes a carreras científicas y tecnológicas que incluyen, en su primer año, asignaturas obligatorias de Matemática y Física con contenidos y criterios de evaluación comparables a lo largo del período analizado. Se consideran cohortes de ingreso consecutivas durante varias décadas, lo que habilita un análisis longitudinal de alta resolución temporal.

## 3.3 Fuentes de datos y variables

Los datos utilizados provienen de registros administrativos institucionales, que incluyen información sobre inscripciones, resultados académicos y condición de aprobación de las asignaturas analizadas. Este tipo de fuentes presenta ventajas significativas para estudios longitudinales extensos, al ofrecer cobertura completa de la población estudiantil y consistencia en los criterios de registro (OECD, 2019).

A partir de estos registros se construyeron las siguientes variables principales:

- **Probabilidad de éxito académico**: definida como la proporción de estudiantes que aprueban la asignatura en el período considerado, ya sea por promoción o por aprobación del examen final.

- **Probabilidad de no intento (fracaso pasivo)**: proporción de estudiantes inscriptos que no registran intentos efectivos de aprobación, indicador asociado a deserción temprana o desvinculación académica (Harackiewicz et al., 2016).

- **Brecha de éxito por tipo de escuela secundaria**: diferencia en la probabilidad de éxito entre estudiantes provenientes de instituciones secundarias privadas y públicas.

Las variables se calcularon de manera agregada por cohorte de ingreso, lo que permite el análisis de tendencias intercohortales y la comparación entre subpoblaciones.

### 3.4 Estrategia analítica

La estrategia analítica combina estadística descriptiva longitudinal con modelos de regresión agregada, orientados a identificar tendencias temporales y cambios estructurales en los indicadores de desempeño. En línea con estudios previos, se prioriza la visualización y modelización de trayectorias intercohortales, en lugar de inferencias causales a nivel individual (Faulkner et al., 2010; Henderson et al., 2015).

Para cada cohorte y asignatura se estimaron probabilidades de éxito y de no intento, diferenciadas por tipo de institución secundaria de procedencia. A partir de estas estimaciones se construyeron series temporales y se identificaron períodos de cambio asociados a modificaciones relevantes del contexto educativo, tales como reformas curriculares o transformaciones en el sistema secundario.

Asimismo, se emplearon modelos de regresión lineal y segmentada para evaluar la magnitud y dirección de las tendencias observadas, así como la estabilidad o variabilidad de las brechas entre subpoblaciones. Este enfoque permite distinguir entre fluctuaciones coyunturales y patrones persistentes de largo plazo, aspecto central para la interpretación evaluativa de los resultados (OECD, 2020).

### 3.5 Consideraciones éticas y limitaciones

El estudio utiliza exclusivamente datos administrativos anonimizados, sin acceso a información personal identificable, en conformidad con las normas éticas vigentes para la investigación educativa basada en registros institucionales (COPE, 2019). No se realizaron intervenciones sobre los estudiantes ni se utilizaron datos sensibles.

Entre las principales limitaciones metodológicas se encuentra la imposibilidad de capturar variables individuales no registradas en los sistemas administrativos, tales como motivación, estrategias de estudio o condiciones socioeconómicas detalladas. No obstante, dado el objetivo del trabajo —analizar tendencias estructurales de preparación preuniversitaria— estas limitaciones no invalidan la pertinencia del enfoque adoptado, sino que delimitan su alcance interpretativo.

## 4. Resultados

### 4.1 Tendencias de largo plazo en la probabilidad de éxito

El análisis longitudinal de las cohortes de ingreso evidencia una tendencia no lineal en la probabilidad de aprobación temprana en las asignaturas sensor de Matemática y Física. En ambas disciplinas se observa un período inicial de incremento progresivo de las tasas de éxito, seguido por una fase de meseta y, posteriormente, por un descenso sostenido en las cohortes más recientes.

Figura 1. Evolución de la probabilidad de aprobación en Matemática por cohorte de ingreso (1980–2019).

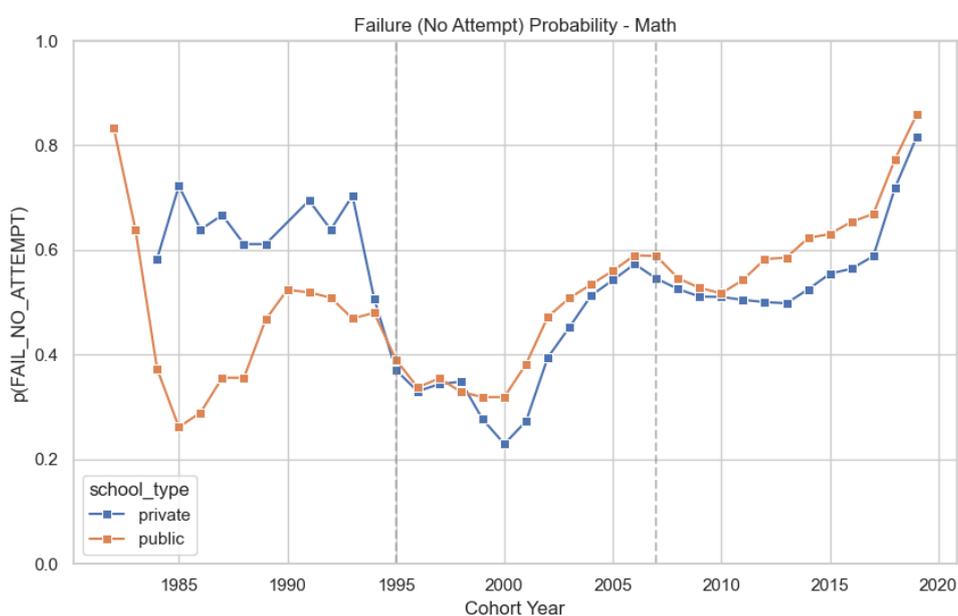

*Nota. La figura presenta la probabilidad de aprobación en el primer intento en la asignatura introductoria de Matemática, estimada a nivel de cohorte de ingreso. Cada punto representa una cohorte anual. Los valores corresponden a resultados agregados y no deben interpretarse como indicadores de desempeño individual.*

En Física, la probabilidad de éxito alcanza valores máximos hacia mediados de la década de 1990, superando el 60 % en algunos años, para luego iniciar un descenso gradual que se acentúa a partir de la década de 2000. En Matemática se detecta un patrón similar, aunque con mayores oscilaciones interanuales y un pico ligeramente posterior, cercano al cambio de milenio. Estos resultados sugieren la presencia de cambios estructurales en la compatibilidad entre la preparación previa de las cohortes y las exigencias académicas iniciales, más que fluctuaciones aleatorias o coyunturales.

Figura 2. Evolución de la probabilidad de aprobación en Física por cohorte de ingreso (1980–2019)

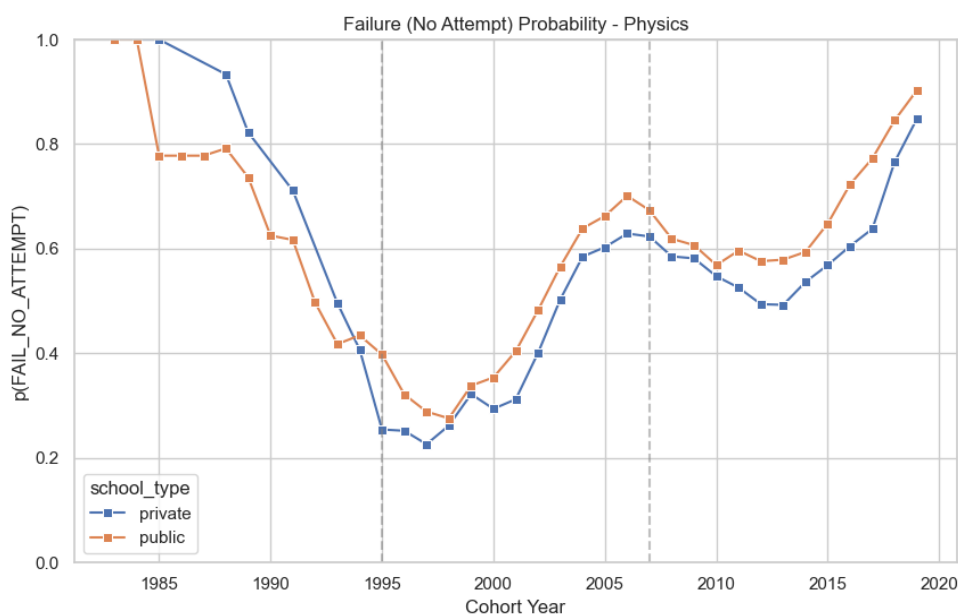

*Nota. La figura muestra la probabilidad de aprobación en el primer intento en la asignatura introductoria de Física, calculada a nivel de cohorte. Los resultados reflejan tendencias longitudinales en el desempeño académico temprano.*

Este tipo de dinámicas de largo plazo ha sido reportado en estudios internacionales que analizan transiciones a la educación superior en contextos masivos, donde las variaciones en el desempeño temprano reflejan transformaciones acumulativas en los sistemas educativos de origen y destino (Faulkner et al., 2010; OECD, 2020).

**4.2 Diferencias por tipo de escuela secundaria**

Al desagregar los resultados según el tipo de institución secundaria de procedencia, se identifica una brecha sistemática en la probabilidad de éxito entre estudiantes provenientes de escuelas privadas y públicas. Dicha brecha es observable tanto en

Matemática como en Física y se mantiene relativamente estable a lo largo del período analizado, aunque con variaciones en su magnitud.

Figura 3. Brecha de aprobación entre estudiantes de escuelas privadas y públicas en Matemática (1980–2019)

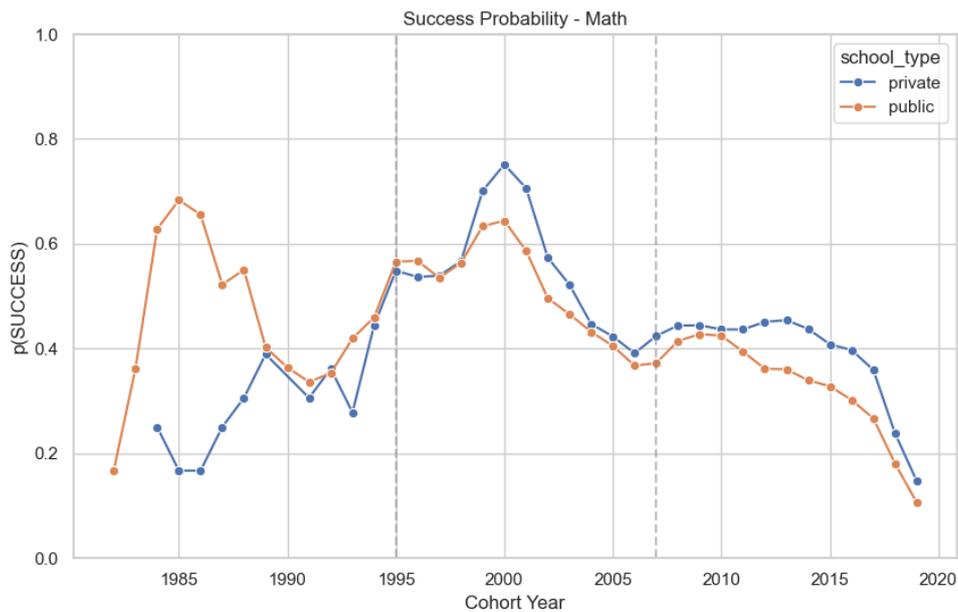

*Nota. La brecha se define como la diferencia en la probabilidad de aprobación en el primer intento entre estudiantes provenientes de escuelas secundarias privadas y públicas. Valores positivos indican una mayor probabilidad de aprobación para el grupo de procedencia privada.*

En Física, la brecha de éxito es reducida o incluso negativa en las cohortes más antiguas, pero se vuelve consistentemente positiva a partir de mediados de la década de 1990, estabilizándose en valores moderados en los años posteriores. En Matemática, la brecha muestra una mayor amplitud en las primeras cohortes, con diferencias marcadas a favor del sector privado, para luego reducirse y estabilizarse en niveles más acotados.

Figura 4. Brecha de aprobación entre estudiantes de escuelas privadas y públicas en Física (1980–2019)

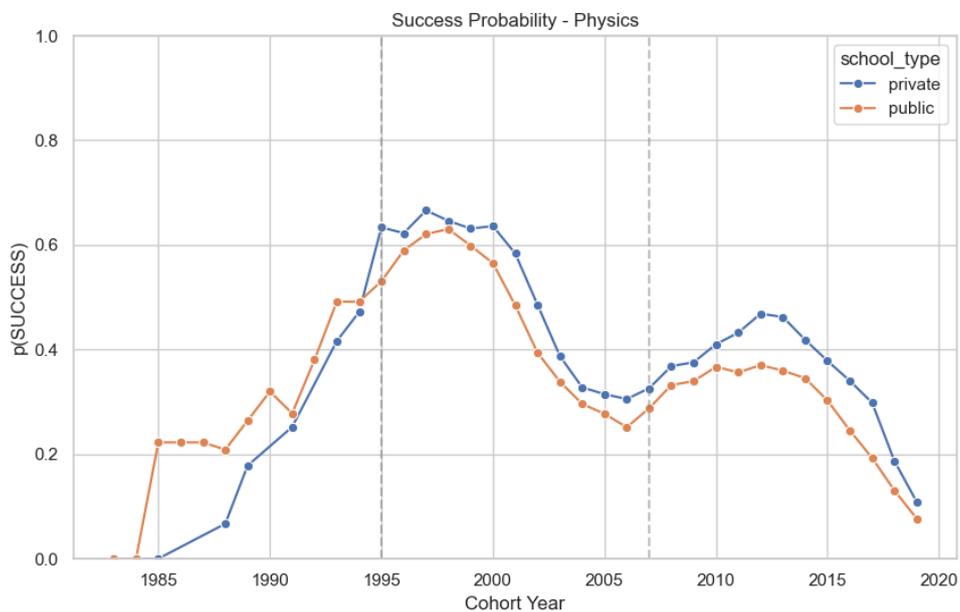

*Nota. La figura representa la evolución longitudinal de la diferencia en la probabilidad de aprobación en Física entre estudiantes de escuelas privadas y públicas. La estabilidad relativa de la brecha sugiere desigualdades persistentes sin divergencia creciente entre ambos grupos.*

Estos patrones indican que, si bien existen diferencias persistentes asociadas al tipo de escuela secundaria, no se observa una divergencia creciente entre ambos grupos. Este hallazgo resulta consistente con investigaciones previas que señalan la persistencia de desigualdades estructurales en el acceso al capital académico, pero también su relativa estabilidad en sistemas de educación superior de acceso abierto (Heublein & Wolter, 2011; OECD, 2019).

**4.3 Dinámicas de no intento y desvinculación temprana**

Uno de los resultados más robustos del análisis es el incremento sostenido de la probabilidad de no intento —definida como la inscripción sin registro de evaluación efectiva— en ambas asignaturas sensor. Este fenómeno se intensifica en las cohortes más recientes y afecta tanto a estudiantes de escuelas públicas como privadas, aunque con niveles absolutos más elevados en el primer grupo.

Figura 5. Evolución de la probabilidad de no intento en Matemática por cohorte de ingreso (1980–2019)

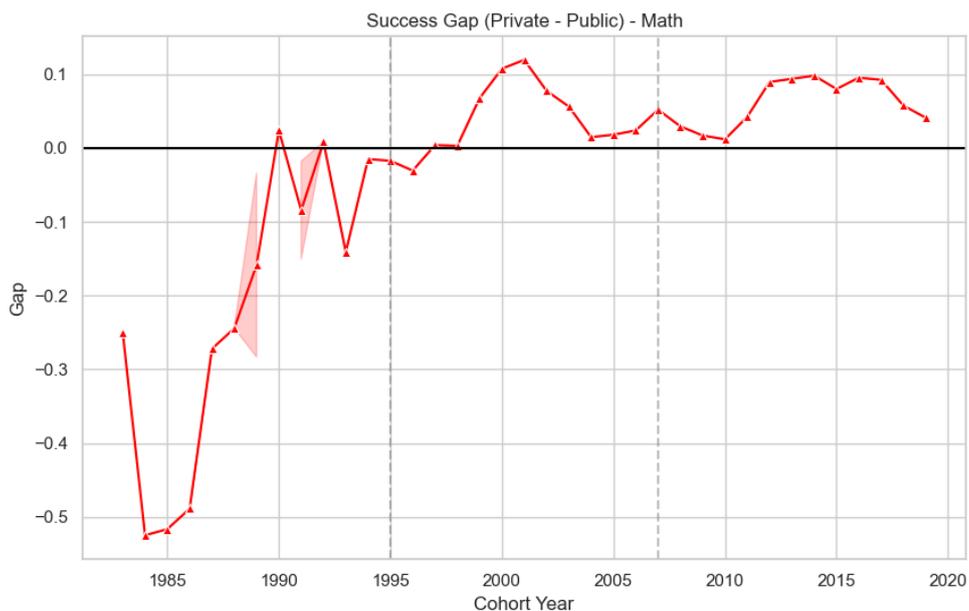

*Nota. La probabilidad de no intento se define como la proporción de estudiantes inscriptos que no registran participación evaluativa efectiva en el período considerado. Este indicador capta formas de desvinculación académica temprana a nivel de cohorte.*

Figura 6. Evolución de la probabilidad de no intento en Física por cohorte de ingreso (1980–2019)

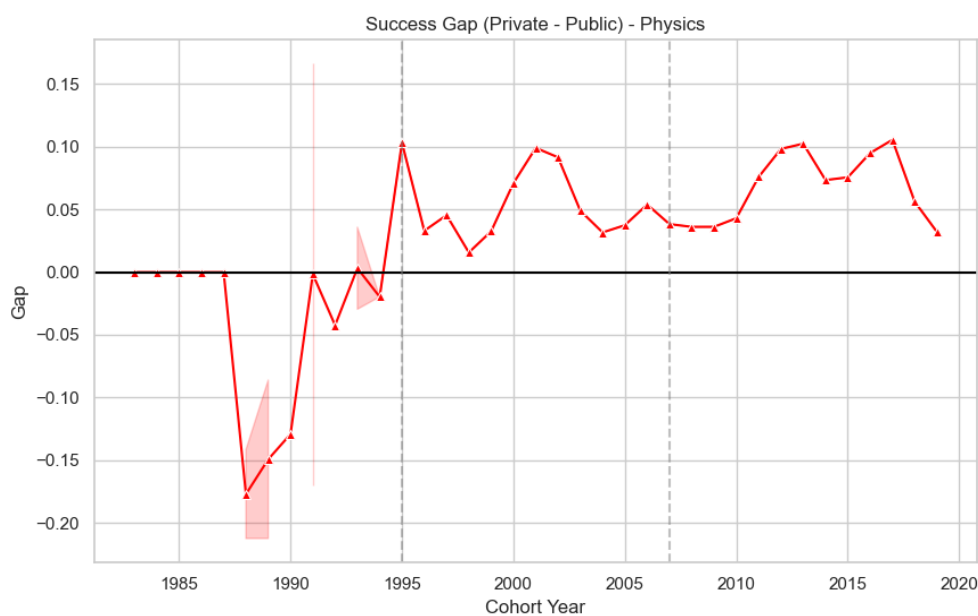

*Nota. La figura muestra la dinámica longitudinal del no intento en la asignatura introductoria de Física. El aumento sostenido del indicador en las cohortes más recientes sugiere cambios estructurales en los patrones de vinculación académica inicial.*

En Física, la probabilidad de no intento disminuye de manera pronunciada hasta mediados de la década de 1990, coincidiendo con el período de mayor éxito académico, y luego inicia un ascenso progresivo que supera el 70 % en las cohortes finales del período analizado. En Matemática se observa un comportamiento análogo, con un mínimo cercano al cambio de milenio y un aumento posterior marcado.

La expansión del no intento sugiere una transformación en los patrones de vinculación académica temprana, en línea con estudios que identifican el abandono temprano y la desactivación académica como fenómenos crecientes en contextos universitarios masivos (Harackiewicz et al., 2016; Henderson et al., 2015). Desde una perspectiva evaluativa, este indicador resulta particularmente relevante, ya que capta formas de incompatibilidad estructural que no se manifiestan únicamente como fracaso académico explícito.

## 4.4 Síntesis de patrones estructurales

Considerados en conjunto, los resultados revelan tres patrones estructurales principales:

(a) una disminución de la probabilidad de éxito temprano en las cohortes más recientes.

(b) una estabilidad relativa —aunque persistente— de las brechas entre estudiantes de escuelas públicas y privadas.

(c) un aumento pronunciado de la desvinculación temprana, expresada en el no intento de evaluación.

Estos patrones no deben interpretarse como indicadores de capacidades individuales, sino como señales agregadas de la relación entre la preparación previa de las cohortes y las demandas académicas iniciales. En este sentido, los cursos sensor funcionan como instrumentos evaluativos indirectos de dicha compatibilidad, permitiendo identificar cambios de régimen en el sistema educativo a lo largo del tiempo, tal como proponen enfoques comparativos de evaluación longitudinal en educación superior (OECD, 2020).

Tabla 1. Coeficientes de los modelos longitudinales por asignatura y tipo de resultado

| Asignatura | Resultado analizado | Término | Coeficiente | IC 95% inferior | IC 95% superior | p-valor |
|---|---|---|---|---|---|---|
| Matemática | Aprobación | Intercepto | 0.259 | 0.163 | 0.356 | < .001 |
| Matemática | Aprobación | Año de cohorte (centrado) | -0.054 | -0.063 | -0.045 | < .001 |
| Matemática | Aprobación | Escuela pública | -0.168 | -0.295 | -0.042 | .009 |
| Matemática | Aprobación | Año × escuela pública | -0.010 | -0.022 | 0.002 | .105 |
| Matemática | No intento | Intercepto | -0.433 | -0.530 | -0.335 | < .001 |
| Matemática | No intento | Año de cohorte (centrado) | 0.054 | 0.046 | 0.063 | < .001 |

| Asignatura | Resultado analizado | Término | Coeficiente | IC 95% inferior | IC 95% superior | p-valor |
|---|---|---|---|---|---|---|
| Matemática | No intento | Escuela pública | 0.134 | 0.006 | 0.261 | .040 |
| Matemática | No intento | Año × escuela pública | 0.010 | -0.001 | 0.022 | .085 |
| Física | Aprobación | Intercepto | -0.057 | -0.159 | 0.044 | .269 |
| Física | Aprobación | Año de cohorte (centrado) | -0.045 | -0.054 | -0.035 | < .001 |
| Física | Aprobación | Escuela pública | -0.171 | -0.305 | -0.037 | .013 |
| Física | Aprobación | Año × escuela pública | -0.015 | -0.027 | -0.002 | .027 |
| Física | No intento | Intercepto | -0.281 | -0.383 | -0.179 | < .001 |
| Física | No intento | Año de cohorte (centrado) | 0.054 | 0.045 | 0.063 | < .001 |
| Física | No intento | Escuela pública | 0.171 | 0.037 | 0.305 | .013 |
| Física | No intento | Año × escuela pública | 0.014 | 0.001 | 0.027 | .027 |

Nota. Los coeficientes corresponden a modelos longitudinales estimados a nivel de cohorte. El término "Año de cohorte (centrado)" representa la tendencia temporal agregada. "Escuela pública" indica la diferencia respecto del grupo de referencia (escuela privada). Los intervalos de confianza son al 95 %. Los resultados se interpretan como asociaciones descriptivas y no implican relaciones causales.

## 5. Discusión

**5.1 Interpretación de las tendencias observadas**

Los resultados presentados muestran transformaciones sostenidas en los patrones de desempeño temprano en asignaturas introductorias de Matemática y Física, entendidas en este estudio como cursos sensor de la compatibilidad entre la preparación preuniversitaria y las exigencias iniciales de la educación superior. La disminución progresiva de la probabilidad de éxito, junto con el aumento del no intento, sugiere una modificación estructural en dicha compatibilidad a nivel de cohorte, más que cambios atribuibles a comportamientos individuales o a factores coyunturales aislados.

Desde una perspectiva de evaluación educativa, estos hallazgos son consistentes con la idea de que los sistemas universitarios de acceso amplio funcionan como dispositivos de detección temprana de tensiones acumuladas en los sistemas de educación previa. En este sentido, los cursos sensor no miden capacidades intrínsecas, sino el grado de alineación funcional entre trayectorias formativas consecutivas (Faulkner et al., 2010; OECD, 2020).

El incremento del no intento reviste especial relevancia interpretativa. A diferencia del fracaso académico explícito, la ausencia de participación evaluativa indica formas de desvinculación temprana que preceden al rendimiento medible y que han sido identificadas en la literatura como señales tempranas de abandono o postergación prolongada (Harackiewicz et al., 2016). Su crecimiento sostenido en el tiempo sugiere cambios en los patrones de engagement académico inicial, más que una simple intensificación de la dificultad curricular.

**5.2 Comparación con la evidencia internacional**

El análisis del presente estudio se alinea parcialmente con tendencias reportadas en investigaciones internacionales sobre la transición a la educación superior en contextos de masificación. Estudios desarrollados en Europa y Oceanía han documentado descensos graduales en el desempeño inicial en disciplinas científicas básicas, así como incrementos en indicadores de abandono temprano o baja participación académica (Faulkner et al., 2010; Henderson et al., 2015).

Asimismo, informes comparativos de la OECD han señalado que, en sistemas de educación superior con acceso amplio, las variaciones longitudinales en el rendimiento temprano suelen reflejar desajustes estructurales entre niveles educativos, más que deterioros homogéneos de la calidad estudiantil (OECD, 2019; OECD, 2020). En este marco, los resultados obtenidos no constituyen una anomalía local, sino que se inscriben en un patrón más amplio observado en distintos sistemas nacionales.

No obstante, la evidencia internacional también muestra heterogeneidad significativa en la magnitud y temporalidad de estos procesos. Algunos sistemas exhiben estabilizaciones tempranas, mientras que otros presentan cambios más abruptos asociados a reformas institucionales específicas. Por ello, los resultados aquí presentados deben interpretarse como parte de un fenómeno de alcance amplio, pero con manifestaciones contextualmente situadas.

### 5.3 Alcance y límites de la inferencia

Es fundamental subrayar que el estudio no permite establecer relaciones causales ni atribuir los cambios observados a factores específicos. La ausencia de variables individuales y contextuales detalladas limita cualquier intento de explicación etiológica. En consecuencia, los resultados deben leerse como descripciones estructurales de largo plazo, útiles para la evaluación del sistema, pero no como diagnósticos exhaustivos de sus causas subyacentes.

Del mismo modo, la persistencia de brechas entre estudiantes de escuelas públicas y privadas no debe interpretarse como evidencia de una divergencia creciente o de una polarización educativa. La relativa estabilidad de dichas brechas a lo largo del tiempo sugiere la presencia de desigualdades estructurales persistentes, pero no necesariamente en expansión, en línea con lo reportado por estudios comparativos sobre equidad en educación superior (Heublein & Wolter, 2011; OECD, 2019).

Finalmente, el carácter institucionalmente situado del estudio impone cautela en la generalización de los resultados. Si bien los patrones identificados dialogan con tendencias internacionales, su interpretación debe anclarse en las características específicas del sistema analizado, particularmente en lo relativo al acceso irrestricto y a la estructura curricular de las carreras consideradas.

## 6. Conclusiones

### 6.1 Principales hallazgos

Este estudio analizó tendencias de largo plazo en la preparación preuniversitaria a partir del desempeño temprano en asignaturas introductorias de Matemática y Física, conceptualizadas como cursos sensor de la compatibilidad entre la formación secundaria y las demandas iniciales de la educación superior. A partir del análisis de múltiples cohortes consecutivas, los resultados evidencian tres patrones robustos: una disminución sostenida de la probabilidad de éxito temprano, un incremento marcado de la no participación evaluativa y la persistencia de brechas moderadas entre estudiantes provenientes de escuelas secundarias públicas y privadas.

Estos hallazgos describen transformaciones estructurales a nivel de cohorte y no deben interpretarse como indicadores de cambios en las capacidades individuales de los estudiantes. En coherencia con la literatura internacional, los cursos iniciales operan aquí como instrumentos de observación sistémica, capaces de revelar tensiones acumuladas entre niveles educativos consecutivos (Faulkner et al., 2010; OECD, 2020).

## 6.2 Qué puede afirmarse y qué no

Los resultados permiten afirmar que la compatibilidad funcional entre la preparación preuniversitaria promedio de las cohortes entrantes y las exigencias académicas iniciales ha experimentado cambios significativos a lo largo del período analizado. En particular, el crecimiento de la no participación evaluativa sugiere una modificación en los patrones de vinculación académica temprana, con implicancias relevantes para la evaluación del sistema universitario en contextos de acceso amplio.

No obstante, el estudio no permite establecer relaciones causales ni identificar factores explicativos específicos de las tendencias observadas. Tampoco es posible atribuir los cambios detectados a políticas educativas concretas, transformaciones tecnológicas o características individuales de los estudiantes. Cualquier interpretación en ese sentido excedería el alcance metodológico del trabajo y contravendría el enfoque evaluativo adoptado (OECD, 2019; Harackiewicz et al., 2016).

## 6.3 Aporte al campo de la evaluación educativa

Desde la perspectiva de la evaluación educativa, este trabajo contribuye al debate metodológico sobre el uso de indicadores tempranos y longitudinales para el análisis del funcionamiento de los sistemas de educación superior. El empleo de cursos sensor y de métricas agregadas por cohorte ofrece una estrategia replicable para el monitoreo de la transición secundaria–universidad, particularmente en sistemas caracterizados por la masificación del acceso y la heterogeneidad de trayectorias previas (Heublein & Wolter, 2011).

Asimismo, el estudio aporta evidencia empírica de largo plazo desde un contexto latinoamericano, un ámbito menos documentado longitudinalmente en la literatura internacional sobre desempeño temprano y abandono universitario, lo que refuerza su relevancia para revistas especializadas en evaluación educativa.

**6.4 Líneas para investigaciones futuras**

Las conclusiones aquí presentadas abren diversas líneas de investigación futura. En primer lugar, resulta pertinente complementar el enfoque agregado con estudios que incorporen variables contextuales y trayectoriales, a fin de explorar posibles mecanismos asociados a las tendencias detectadas. En segundo lugar, comparaciones sistemáticas entre instituciones y países permitirían evaluar el grado de generalización de los patrones observados y su dependencia de configuraciones institucionales específicas.

Finalmente, el uso de indicadores de no intento y desvinculación temprana merece un desarrollo metodológico más profundo dentro del campo de la evaluación educativa, dado su potencial para captar formas tempranas de incompatibilidad académica que no son visibles a través de las métricas tradicionales de rendimiento.

**Referencias**